\documentclass[
  ,draft            
]
  {aipproc}

\layoutstyle{6x9}

\def\slashchar#1{\setbox0=\hbox{$#1$}
   \dimen0=\wd0 \setbox1=\hbox{/} \dimen1=\wd1
   \ifdim\dimen0>\dimen1 \rlap{\hbox to \dimen0{\hfil/\hfil}} #1
   \else  \rlap{\hbox to \dimen1{\hfil$#1$\hfil}} / \fi}
\def\P{\slashchar{P}}

\begin{document}

\title{Meson-Baryon Interactions in Unitarized Chiral Perturbation
Theory\thanks{Presented by E. Ruiz Arriola at the 2nd International
Workshop On Hadron Physics: Effective Theories Of Low-Energy QCD
Coimbra 2002}}

\author{G. Garc{\'i}a Recio}{
address={Departamento de F{\'{\i}}sica Moderna,  
Universidad de Granada. 18071-Granada (Spain) }
}
\author{J. Nieves}
{
address={Departamento de F{\'{\i}}sica Moderna,  
Universidad de Granada. 18071-Granada (Spain) }
}
\author{\underline{E.Ruiz Arriola}}
{
address={Departamento de F{\'{\i}}sica Moderna,  
Universidad de Granada. 18071-Granada (Spain) }
}
\author{M. Vicente Vacas}{
  address={Departamento de F\'{\i}sica Te\'orica and IFIC, 
Centro Mixto Universidad de Valencia-CSIC, 
Ap. Correos 22085, E-46071 Valencia (Spain)}}

\begin{abstract}
Meson-Baryon Interactions can be successfully described using both
Chiral Symmetry and Unitarity. The $s-$wave meson-baryon scattering
amplitude is analyzed in a Bethe-Salpeter coupled channel formalism
incorporating Chiral Symmetry in the potential. Two body coupled
channel unitarity is exactly preserved.  The needed two particle
irreducible matrix amplitude is taken from lowest order Chiral
Perturbation Theory in a relativistic formalism. Off-shell behavior
is parameterized in terms of low energy constants. The relation to the
heavy baryon limit is discussed. The position of the complex poles in
the second Riemann sheet of the scattering amplitude determine masses
and widths baryonic resonances of the $N(1535)$, $N(1670)$, $\Lambda
(1405)$ and $\Lambda(1670)$ resonances which compare well with
accepted numbers.
\end{abstract}

\maketitle


\section{Introduction}
\label{intro}
The existence of baryon resonances in meson-baryon reactions is a
non-perturbative feature of QCD at intermediate energies. Although
Relativistic Invariance, Crossing Symmetry, Unitarity and Chiral
Symmetry undoubtedly provide powerful constraints, the usual high and
low energy simplifications do not directly apply here just because
there is no obvious expansion parameter and some special methods and
approximations have to be developed.

The nature of baryonic resonances is not fully unambiguous since their
lifetime, $\tau_R$, is very short and hence they cannot be observed
decaying into final products. The standard and accepted definition is
that resonances correspond to poles in the complex CM energy plane in
unphysical sheets, $ -2 {\rm Im} \sqrt{s_R}= \Gamma_R = 1/\tau_R $
complying with causality requirements. It is remarkable that although
this definition has long been known, it has only been very recently
incorporated explicitly in the Particle Data~\cite{pdg02}. Other
definitions (zeros of the K-matrix, phase shift passing through
$90^o$, maximum of cross section, etc.), although simpler to work
with, coincide with the former one only in the limit $\Gamma_R \to
0$. For non-relativistic potential scattering this limit produces the
exponential law and corresponds to the picture that a stable particle
tunnels through the barrier to the continuum.  Nevertheless, one
should pay attention to the fact that going to the complex $s-$plane
in terms of data on the real axis may be a model dependent operation,
particularly for not very sharp resonances. Conversely, going from the
pole in the unphysical sheet to the real scattering line is also not
uniquely defined, and the energy dependence may be model dependent as
well. This ambiguity is enhanced in the case where there are several
open channels, because there are many ways of parameterizing a
multichannel scattering amplitude with a given analytical
structure~\cite{BK82}. The situation is aggravated by the fact that
very often the analysis of data in terms of partial wave amplitudes
are not provided with error estimates.

In the quark model baryon resonances are naturally interpreted as
bound state composites of three valence quarks, and their widths are
computed as matrix elements of hadronic transition operators between
the bound and continuum states~\cite{AM00}. This approach corresponds
to the picture of unstable particles weakly coupled to a continuum. To
improve on this, the scattering problem has to be solved. The
information entering such a problem is baryon masses and baryon-meson
form factors; the only specific reminder of the underlying quark
degrees of freedom has to do with these form factors. Although quark
models are very fruitful and provide a lot insight into the problem
they suffer at least from two deficiencies regarding the description
of meson-baryon scattering. Firstly, assuming that masses of hadrons
come out properly as a result of a quark model calculation, there is
still the possibility that it fails when describing the scattering
process. This could only be interpreted as a failure of the particular
set of interactions or approximations used to solve the quark model,
but does not prevent from finding another quark model providing a
better description. In the second place, at the hadronic level
symmetries, such as relativistic invariance and chiral symmetry, which
are known to work well, are difficult to impose at the hadronic level
if hadrons are described as composites.

An alternative and not necessarily incompatible (although perhaps more
economical) point of view to the quark model calculations is to
formulate the whole problem directly in terms of hadronic degrees of
freedom.  This allows to impose all known information, symmetries in
particular, from the beginning, and offers the possibility to falsify
all unknown information. It is of course impossible to do this at all
energies, but it may be achieved at low energies, in terms of a finite
and countable number of unknown parameters, the so-called low energy
constants.  This is the point of view of Chiral Perturbation
Theory~\cite{Gasser:1983yg,Pich95}.  At the level of a relativistic
Lagrangean this corresponds to write down the most general infinite
set of tree level operators compatible with the known
symmetries. Since a tree level Lagrangean only produces real
amplitudes and hence violates unitarity it is necessary to incorporate
quantum (loop) corrections.

In the meson-baryon system there is a problem in Chiral Perturbation
Theory already found long ago~\cite{gss88} because in the standard
dimensional regularization, heavy particles do not decouple. This
result is counter-intuitive because it means that particles with a
very small Compton wavelength propagate.  Two approaches have been
suggested to overcome this difficulty. In Heavy Baryon Chiral
Perturbation Theory (HBChPT)~\cite{JM91,BK92} one takes the
non-relativistic limit first and then proceeds in dimensional
regularization. The heavy particle decoupling is explicitly built in,
but relativistic invariance is not manifest at any step of the
calculation. Within this framework elastic $\pi N$ scattering has been
studied to third~\cite{Mo98,fms98} and fourth order~\cite{fm00} in the
chiral expansion and also $\bar K N $ elastic
scattering~\cite{Ka01}. A more recent proposal keeps relativistic
invariance explicitly at any step of the calculation but introduces a
new so-called infrared regularization which complies with decoupling
in the heavy particle baryon~\cite{bl01} and allows a satisfactory
description of $\pi N$ elastic scattering~\cite{Becher:2001hv}. In
either case, although crossing is exact at any order of the expansion
unitarity is only built in perturbatively. The need for unitarization
in the $S=-1$ channel becomes obvious after the work of
Ref.~\cite{Ka01} where it is shown that HBChPT to one loop fails
completely in the $\bar K N $ channel already at threshold due to
nearby subthreshold $\Lambda (1405) $-resonance.

The Bethe-Salpeter equation (BSE) provides the framework beyond
perturbation theory to treat the relativistic two body problem from a
Quantum Field Theory point of view. This approach allows to treat the
scattering problem preserving exact unitarity. In practical
applications, however, the number of particles is kept fixed and other
approximations are done, violating crossing symmetry.  At the level of
partial waves unitarity implies a right cut discontinuity while
crossing generates a left cut for the scattering amplitude, but in the
scattering region one expects the energy dependence to by mainly
determined by the right cut.

The $s-$wave meson-baryon scattering incorporating chiral symmetry and
unitarization for several open channels has been studied in previous
works~\cite{KSW95,OR98,Kr98,OM01,Ke01,ORB02,LK02}.  The purpose of the
present contribution is to give a brief overview of {\it a possible}
approximation scheme for meson-baryon scattering based on the
BSE. This is the natural extension of work previously done for
meson-meson scattering~\cite{EJ99} to the meson-baryon
system~\cite{JE01} for heavy baryons and in a relativistic
formulation~\cite{JE01b,CJ02a,CJ02}.

\section{S-Wave Meson-Baryon Scattering}
The coupled channel scattering amplitude for the baryon-meson process
in given isospin channel $I$ is given by
\begin{equation} 
\left( T_P \right)_{BA} = \bar u_B ( P-k', s_B) t_P (k,k') u_A
 (P-k,s_A)\label{eq:deftpeque}
\end{equation} 
Here, $u_A (P-k, s_A)$ and $u_B (P-k', s_B)$ are baryon Dirac spinors
normalized as ${\bar u }u = 2M$, $P$ is the conserved total CM four
momentum, $P^2 =s$, and $t_P (k,k') $ is a matrix in the Dirac and
coupled channel spaces. Details on normalizations and definitions of
the amplitudes can be seen in Ref.~\cite{JE01b}.  

On the mass shell and using the equations of motion for the free Dirac
spinors $( \slashchar{P} - \slashchar{k} - M_A ) u_A (P-k)=0 $ and its
transposed $ {\bar u}_A (P-k) ( \slashchar{P} - \slashchar{k} - M_A )
=0 $ the parity and Lorentz invariant amplitude $t_P$ relevant
$s-$wave scattering can be simply written as a matrix function in
coupled channel space of $\P$ with $P$ the total CM momentum
\begin{equation}
t_P (k,k') |_{\rm on-shell} = t (\P) 
\label{eq:amp-dirac}
\end{equation} 
In terms of the matrix $t (\P)$ defined in Eq.~(\ref{eq:amp-dirac}),
the $s-$wave coupled-channel matrix, $f_0^\frac12 (s)$, is simply
given by:
\begin{eqnarray}
\left[ f_0^\frac12 (s) \right]_{B \leftarrow A}&=& -\frac{1}{8\pi\sqrt s}
\sqrt{\frac{|\vec{k}_B|}{|\vec{k}_A|}} \sqrt{E_B + M_B}\sqrt{E_A +
M_A} 
\left[t(\sqrt{s} )  \right]_{BA}
\label{eq:deff0}
\end{eqnarray}
where the CM three--momentum moduli read 
\begin{eqnarray}
|\vec{k}_i| &=& \frac{\lambda^\frac12 (s,M_i,m_i)}{ 2\sqrt {s}}\qquad i=A,B
\end{eqnarray}
with $\lambda(x,y,z) = x^2+y^2+z^2 -2xy-2xz-2yz$ and $E_{A,B}$ the
baryon CM energies. The phase of the matrix $T_P$ is such that the
relation between the diagonal elements ($A=B$) in the coupled channel
space of $f_0^\frac12 (s)$ and the in-elasticities ($\eta$) and
phase-shifts ($\delta$) is the usual one,

\begin{eqnarray} 
\left [ f_0^\frac12 (s)\right]_{AA}  = {1\over 2 {\rm i} |\vec{k}_A| } 
\Big( \eta_A (s) e^{2
{\rm i} \delta_A (s)} - 1 \Big) 
\label{eq:delta0} 
\end{eqnarray}
Hence, the optical theorem reads, for $s \ge (M_A+m_A)^2$,
\begin{eqnarray}
\frac{4\pi}{|\vec{k}_A|} {\rm Im} \left [ f_0^\frac12 (s)\right]_{AA}  &=&
\sum_B \sigma_{B \leftarrow A} = 4\pi \sum_B \left |\left[  f_0^\frac12 (s) \right]_{BA}\right|^2 =  \sigma_{AA} + \frac{\pi}{|\vec{k}_A|^2}
\left(1-\eta_A^2\right) 
\label{eq:optical} 
\end{eqnarray} 
where in the right hand side only open  channels contribute.

\section{Chiral Baryon-Meson Lagrangian}

At lowest order in the chiral expansion the chiral baryon meson
Lagrangian contains kinetic and mass baryon pieces and meson-baryon
interaction terms and is given by~\cite{Pich95}
\begin{eqnarray}
{\cal L}_1 = {\rm Tr} \left\{ \bar{B} \left( {\rm i}
\slashchar{\nabla} - M_B \right) B \right\} + 
\frac{1}{2} \, {\cal D} \, {\rm Tr} \left\{ 
\bar{B} \gamma^\mu \gamma_5 \left\{ u_\mu , B
\right\} \right\} + \frac{1}{2} \, {\cal F} \, {\rm Tr} \left\{ 
\bar{B} \gamma^\mu \gamma_5 [ u_\mu ,B] \right\}  \, ,
\label{LB1}
\end{eqnarray}
The meson kinetic and mass pieces and the baryon mass chiral
corrections are second order and read
\begin{eqnarray}
{\cal L}_2 &=& {f^2 \over 4} {\rm Tr} \left\{ u_\mu^\dagger u^\mu + 
(U^\dagger \chi + \chi^\dagger U ) \right\} \nonumber \\ &-&
b_0 {\rm Tr} ( \chi_+ ) {\rm Tr} (\bar B B) - b_1 {\rm Tr} ( \bar B
\chi_+ B ) - b_2 {\rm Tr} ( \bar B B \chi_+ )
\label{LB2}
\end{eqnarray}
where ``Tr'' stands for the trace in $SU(3)$. In addition,
\begin{eqnarray}
\nabla_{\mu} B &=& \partial_{\mu} B + \frac{1}{2}\, [ \, u^\dagger
\partial_{\mu} u + u \partial_{\mu} u^\dagger \, , \, B \, ] \,
,\nonumber \\
\label{LB1_exp}
U = u^2 &=& e^{ {\rm i} \sqrt{2} \Phi / f } \, , \qquad u_{\mu} = {\rm
i} u ^\dagger \partial_{\mu} U u^\dagger \, \nonumber \\ \chi_+ &=&
u^\dagger \chi u^\dagger + u \chi^\dagger u \, , \qquad \chi = 2 B_0
{\cal M}
\end{eqnarray} 
$M_B$ is the common mass of the baryon octet, due to spontaneous
chiral symmetry breaking for massless quarks. The $SU(3)$ coupling
constants which are determined by semileptonic decays of hyperons are
${\cal F} \sim 0.46$, ${\cal D} \sim 0.79$ (${\cal F}+{\cal D} = g_{A}
= 1.25$).  The constants $B_0$ and $f$ (pion weak decay constant in
the chiral limit) are not determined by the symmetry. The current
quark mass matrix is ${\cal M}={\rm Diag}(m_u,m_d,m_s)$. The
parameters $b_0$, $b_1$ and $b_2$ are coupling constants with
dimension of an inverse mass. The values of $b_1$ and $b_2$ can be
determined from baryon mass splittings, whereas $b_0$ gives an overall
contribution to the octet baryon mass $M_B$. Neglecting
octet-singlet mixing, the $SU(3)$ matrices for the meson and the
baryon octet are written in terms of the meson and baryon spinor
fields respectively and are given by
\begin{eqnarray}
	\Phi = \left( \matrix{ \frac{1}{\sqrt{2}} \pi^0 +
	\frac{1}{\sqrt{6}} \eta & \pi^+ & K^+  \cr  \pi^- & -
	\frac{1}{\sqrt{2}} \pi^0 + \frac{1}{\sqrt{6}} \eta & K^0  \cr 
	K^- & \bar{K}^0 & - \frac{2}{\sqrt{6}} \eta }
	\right) \, ,
\end{eqnarray}
and 
\begin{eqnarray}
	B =
	\left( \matrix{ 
	\frac{1}{\sqrt{2}} \Sigma^0 + \frac{1}{\sqrt{6}} \Lambda &
		\Sigma^+ & p \cr 
		\Sigma^- & - \frac{1}{\sqrt{2}} \Sigma^0 
		+ \frac{1}{\sqrt{6}} \Lambda & n \cr 
		\Xi^- & \Xi^0 & - \frac{2}{\sqrt{6}} \Lambda } 
	\right) \, .
\end{eqnarray}
respectively. The $MB \to MB$ vertex obtained from the former
Lagrangian reads
\begin{equation}
\label{Lmbmb}
	{\cal L}_{MB \to MB} = \frac{\rm i}{4 f^2}{\rm Tr} \left\{
 \bar{B} \gamma^\mu \left[ \, [ \, \Phi \, , \, \partial_\mu \Phi \, ]
 \, , \, B \, \right] \, \right\} \, .
\end{equation}
Assuming isospin conservation, the scattering amplitude ( convention
$-iT_{MB\to MB} =+i{\cal L}_{MB\to MB}$) in the Dirac spinor basis,
at lowest order is given by
\begin{equation}
t_P^{(1)} (k,k') = {D \over f^2} ( \slashchar{k}+\slashchar{k}' )
\label{eq:lowest} 
\end{equation}
where $k$ and $k'$ are incoming and outgoing meson momenta and $D$ a
coupled-channel matrix. On the mass shell one can use Dirac's equation
and gets 
\begin{equation}
t^{(1)}(\P) 
\equiv t_P^{(1)} (k,k')|_{\rm on-shell}  = {1\over f^2} \left\{ \P - \hat M , D \right\} 
\label{eq:lowest_onshell} 
\end{equation} 
which depends only on the CM momentum $\P$. Obviously, the s-wave
scattering amplitude $f_{BA}(s)$, defined through Eq.~(\ref{eq:deff0})
is real and hence cannot satisfy the optical theorem,
Eq.~(\ref{eq:optical}). Thus, there is need for unitarization.

To take into account the some chiral symmetry breaking effects
physical mass splittings and different decay constants must be
accounted for. This can be easily accomplished through the
prescription
\begin{eqnarray}
D/f^2 \to \hat f^{-1} D \hat f^{-1}  \label{eq:f-presc}
\end{eqnarray}  
where $\hat f$ is a diagonal matrix in the coupled channel space.  We
will use the $D/f^2$ notation throughout, meaning
Eq.~(\ref{eq:f-presc}) in practice.
\begin{figure}[ttb]
   \includegraphics[height=.1\textheight]{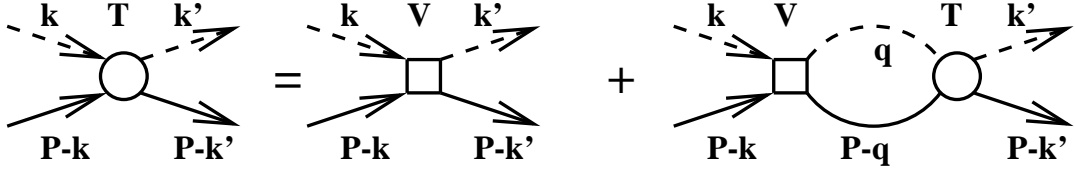} 
  \caption{Diagrammatic representation of the Bethe-Salpeter Equation
  for Meson-Baryon Scattering. $k$ is the meson momentum and $P-k$
  the Baryon momentum. $s=P^2$.}
\label{fig:bse} 
\end{figure}

\section{Bethe-Salpeter Equation} 
\subsection{Two Particle Unitarity} 
To evaluate the meson-baryon scattering amplitude $t_P$ we use the BSE
 (see Fig.~(\ref{fig:bse})) in a given isospin and strangeness
 channel,
\begin{eqnarray} 
t_P (k,k') = v_P (k,k') +  {\rm i}  \int { d^4 q \over
(2\pi)^4 } t_P (q,k') \Delta(q) S(P-q) v_P (k,q) \label{eq:bse}
\end{eqnarray}
where $t_P( k,k')$ is the scattering amplitude defined in
Eq.~(\ref{eq:deftpeque}), $v_P(k,k')$ the two particle irreducible
Green's function (or {\it potential} ), and $ S(P-q)$ and $\Delta (q)
$ the baryon and meson exact propagators respectively. The above
equation turns out to be a matrix one, both in the coupled channel and
Dirac spaces. For any choice of the {\it potential} $v_P(k,k')$, the
resulting scattering amplitude $t_P( k,k')$ fulfills the coupled
channel unitarity condition,
\begin{eqnarray}
t_P(k,k') - {\bar t}_P(k',k) = {\rm i} \int\frac{d^4
q}{(2\pi)^4} t_P(q,k')\, {\rm Disc} \left[ \Delta (q) S(P-q) \right] {\bar t}_P(q,k)
\label{eq:off-uni}
\end{eqnarray} 
where $ {\bar t}_P(k,p) = \gamma_0 {t}_P^\dagger (k,p) \gamma_0 $ and
$ {t}_P^\dagger (k,p) $ is the total adjoint in the Dirac and coupled
channel spaces, including also the discontinuity change $s + {\rm i}
\epsilon \to s - {\rm i} \epsilon $). The two particle discontinuity
is given by Cutkosky's rules, 
\begin{eqnarray}
{\rm Disc} \left[ \Delta (q) S(P-q) \right]= (-2\pi {\rm i} )^2
\delta^+ \left[ q^2-{\hat m}^2 \right] \left( \slashchar{P} -
\slashchar{q} + {\hat M} \right) \delta^+ \left[ (P-q)^2-{\hat M}^2
\right]
\end{eqnarray} 
$\hat m$ and $\hat M$ are meson and baryon (diagonal) mass matrices
respectively and $\delta^+(p^2 - m^2) = \Theta(p^0) \delta(p^2-m^2)$
is the on-shell condition. If the on-shell amplitude depends on $\P$
only, one gets the very simple relation discontinuity equation 
for the inverse amplitude, 
\begin{equation}
{\rm Disc}\, {t(\P)^{-1} } = - {\rm Disc} \, J(\P ) 
\label{eq:on-uni}
\end{equation}  
where the quadratic and logarithmically divergent one-loop integral
\begin{equation}
J(\P ) = {\rm i} \int { d^4 q \over (2\pi)^4 } {1\over q^2 -
\hat m^2}{1\over \slashchar{P}-\slashchar{q} - \hat M  } 
\end{equation}  
has been introduced. Direct calculation leads to
\begin{eqnarray}
J(\P ) &=& \slashchar{P} \left[ \left( { s-\hat m^2+\hat M^2 \over 2
s} \right) J_0 (s) + {\Delta_{\hat m}-\Delta_{\hat M} \over 2 s }
\right] + \hat M J_0 (s) \label{eq:JP}
\end{eqnarray} 
where the quadratic divergences
\begin{eqnarray} 
\Delta_{\hat m} = {\rm i} \int { d^4 q \over (2\pi)^4 } {1\over q^2 -
\hat m^2} \, ,  \qquad \Delta_{\hat M} = {\rm i} \int { d^4 q \over (2\pi)^4
} {1\over q^2 - \hat M^2} \label{eq:deltafit}
\end{eqnarray}  
and the standard logarithmically divergent one loop function
\begin{eqnarray}
J_0 (s) &=& {\rm i} \int { d^4 q \over (2\pi)^4 } {1\over q^2 - \hat
m^2}{1\over (P-q)^2 - \hat M^2 }= \bar J_0 (s) + \hat J_{mM}  
\label{eq:J0mM}
\end{eqnarray}
have been introduced. Here $\bar J_0 (s)$ is normalized to vanish at
threshold, $s=(m+M)^2$, and $\hat J_{mM} $ is a divergent subtraction
constant. 
\subsection{Lowest Order Solution} 
The BSE requires some input potential and baryon and meson propagators
to be solved. At lowest order of the BSE-based chiral expansion, we
approximate the iterated {\it potential} by the chiral expansion
lowest order meson-baryon amplitudes in the desired strangeness and
isospin channels, and the intermediate particle propagators by the
free ones (which are diagonal in the coupled channel space). From the
meson-baryon chiral Lagrangian, one gets at lowest order
for the {\it potential}
\begin{equation}
v_P (k,k') = t_P^{(1)} (k,k') = {D \over f^2} ( \slashchar{k}+\slashchar{k}' ) 
\end{equation}
The $s-$wave BSE can be solved up to a numerical matrix inversion in
the coupled channel space~\cite{JE01b}. The result for the inverse
on-shell amplitude reads
\begin{eqnarray}
t (\P) ^{-1} +  &=& -J(\P)+{\Delta_{\hat m} \over \P -
\hat M} + \left\{ v (\P) ^{-1} + {1\over f^4} (\P- \hat M) D
{\Delta_{\hat m} \over \P- \hat M } D (\P -\hat M) \right\}^{-1}
\label{eq:t-1dirac}
\end{eqnarray} 
This solution manifestly fulfills the on-shell unitarity condition,
Eq.~(\ref{eq:on-uni}). Notice that if $\Delta_{\hat m}=\Delta_{\hat
M}=0$ the tree level {\it on-shell} potential $v(\P)=t^{(1)} (\P) $
determines the amplitude up to a constant $J_{mM}$. Thus, the
difference to do with off-shell effects which we parameterize in terms
of divergent constants. Assuming a specific cut-off method for the
divergent integrals $J_{mM}$, $\Delta_m$ and $\Delta_M$ embodies
specific correlations among them, but quite generally one expects them
to be uncorrelated.
\subsection{Renormalization} 
To renormalize the amplitude given in Eq.~(\ref{eq:t-1dirac}), we note
that in the spirit of an Effective Field Theory (EFT) all possible
counter-terms should be considered. This can be achieved in our case
in a perturbative manner, making use of the formal expansion of the
bare amplitude $T = V + VG_0V + VG_0VG_0V+ \cdots $, where $G_0$ is
the two particle propagator. Thus, a counter-term series should be
added to the bare amplitude such that the sum of both becomes
finite. At each order in the perturbative expansion, the divergent
part of the counter-term series is completely determined. However, the
finite piece remains arbitrary as long as the used {\it potential} $V$
and the meson and baryon propagators are approximated rather than
being the exact ones. Our renormalization scheme is such that the
renormalized amplitude can be cast, again, as in
Eq.~(\ref{eq:t-1dirac}).  This amounts in practice, to interpret the
previously divergent quantities $\hat J_{mM}$, $\Delta_{\hat m}$ and
$\Delta_{\hat M}$ as 12 renormalized free parameters for the $s-$wave
lowest order amplitude in a given isospin-strangeness channel. These
parameters and therefore the renormalized amplitude can be expressed
in terms of physical (measurable) magnitudes. In principle, they
encode the unknown short distance behavior, in particular the
composite nature of hadrons which becomes relevant when they start
overlapping. In practice it seems convenient to fit them to the
available data, although they might be computed within models.
\subsection{Number of parameters} 
Within the spirit of an effective field theory at the hadronic level,
the number of adjustable parameters should not be smaller than those
allowed by the symmetry; this is the only way both to falsify all
possible theories embodying the same symmetry principles and to make
wider the energy window which is being described. The opposite
situation, i.e. a redundancy of parameters is also not desirable, but
less problematic because it may be detected. The precise number of
unknown parameters is mainly controlled to any order of the
calculation by crossing symmetry. In a unitarized approach, the best
way to avoid this parameter redundancy is to match the unitarized
amplitude to one obtained from a Lagrangian formalism as was
explicitly done for meson-meson scattering~\cite{EJ99}. Unfortunately,
there is no standard one loop ChPT calculation for the meson-baryon
reaction with open channels to compare with. We comment on this
matching below. At present the only practical, but indirect, way to
detect such a parameter redundancy is through a fit to experimental
data if the correlations in some parameters turn out to be very
strong.
\begin{figure}[t]
\includegraphics[height=0.7\textheight]{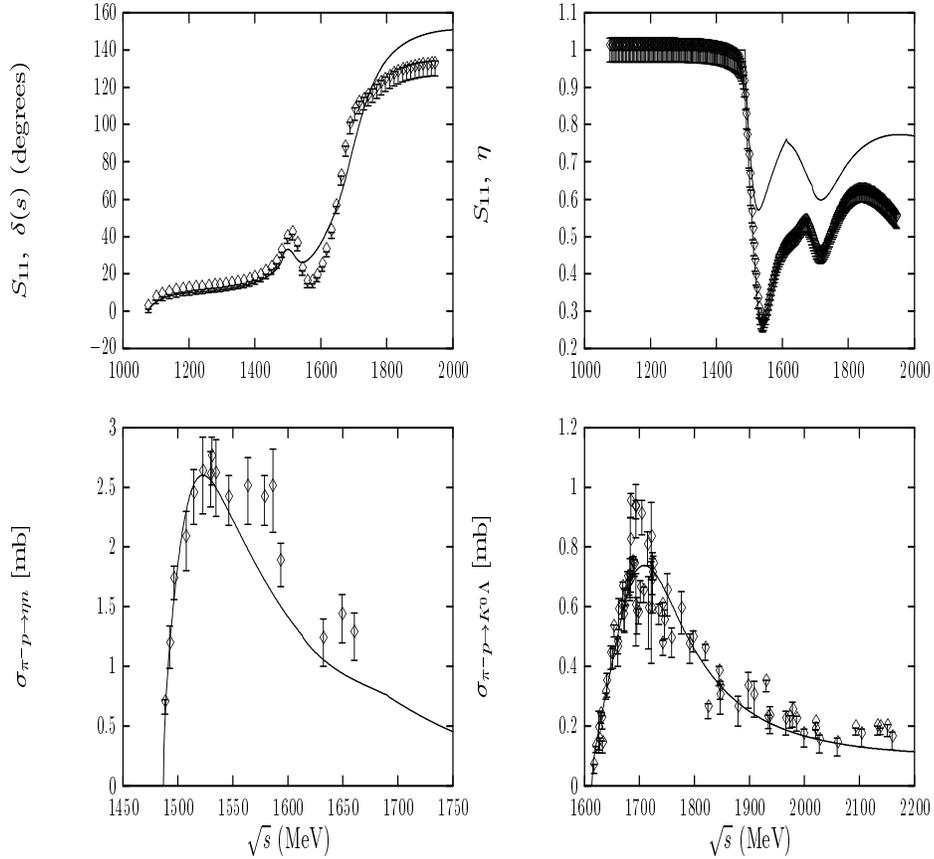}
\caption{ $\pi N $ scattering
BSE results as a function of C.M. energy $\sqrt{s}$. Upper left
figure: $S_{11}$ phase shifts. Upper right figure: inelasticity inqy
the $\pi N$ channel. Lower left figure: $\pi N \to \eta N $ cross
section. Lower right figure: $\pi N \to K \Lambda $ cross section.
Data from Ref.~\cite{AS95}.  (See Ref.~\cite{JE01b} for further
details.)} \label{fig:S=0}
\end{figure}
\subsection{Numerical Results} 
\begin{figure}

\resizebox{.9\textwidth}{!}{
\includegraphics{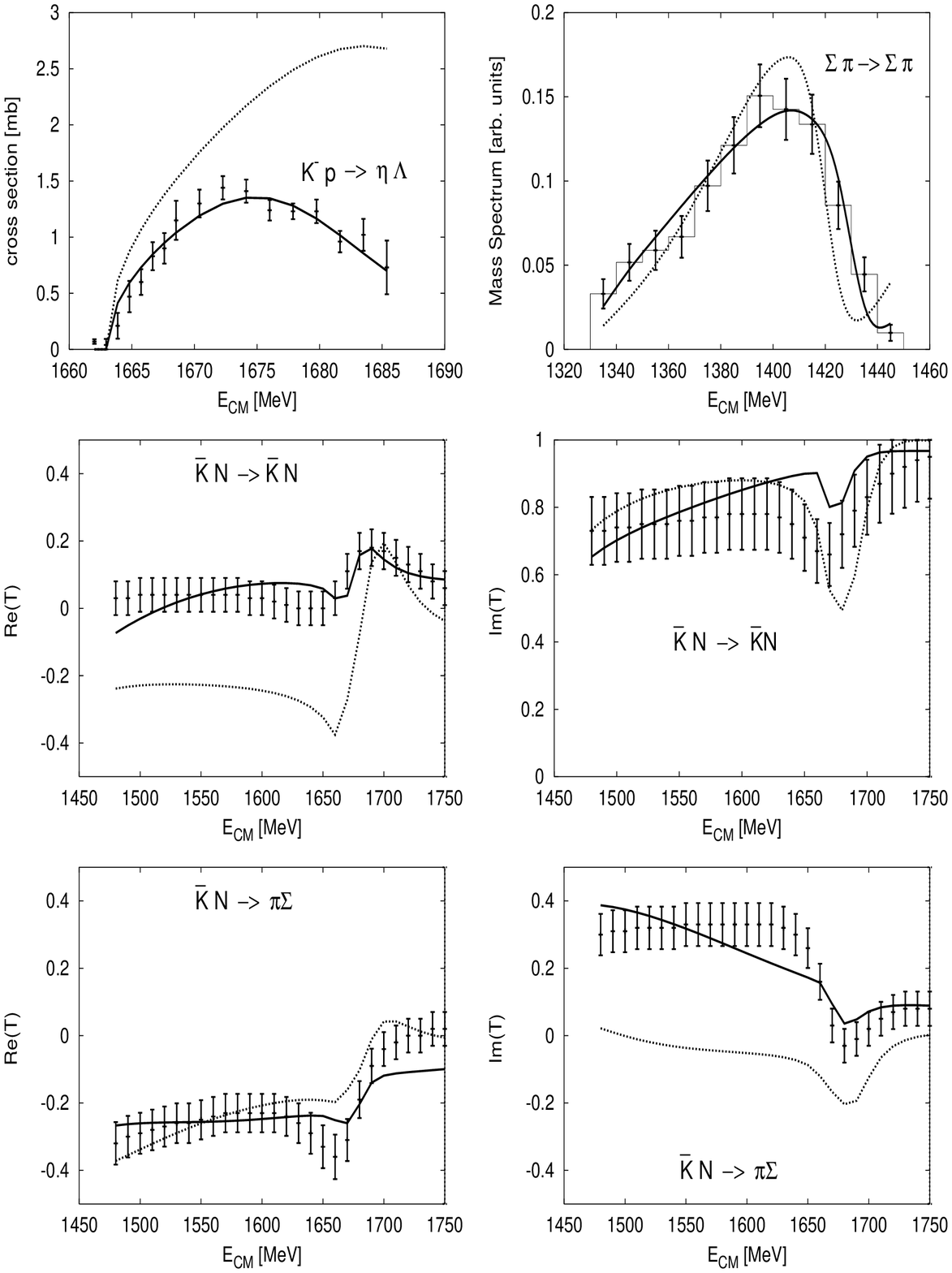}
}
\caption{Best fit results for the Bethe-Salpeter equation in the
$S=-1$, $I=0$ channel (solid lines) compared to
Ref.~\protect{\cite{ORB02}} (doted lines) where $\Delta_m=\Delta_M=0
$. {\bf Upper pannel:} Experimental data for $\pi\Sigma\to\pi\Sigma$
and $K^- p\to \eta\Lambda$ are from
Refs.~\protect\cite{He84}~and~\protect\cite{St01}, respectively. {\bf
Middle pannel:} The real (left panel) and imaginary (right panel)
parts of the $s-$wave $T-$matrix, with normalization specified in the
main text, for elastic $\bar K N \to \bar K N$ process in the $I=0$
isospin channel as functions of the CM energy.  Experimental data are
taken from the analysis of Ref.~\protect\cite{Go77} with the errors
stated in Ref.~\protect\cite{CJ02}. {\bf Lower Pannel:} Same as middle
pannel for the inelastic channel $\bar K N \to \pi \Sigma$.}
\label{fig:S=-1}       
\end{figure}

To finish the presentation, we show some selected numerical results
for the $I=1/2$, $S=0$ and $I=0$, $S=-1$ channels. Details of the
fitting procedure and the relevant experimental data as well as a
thorough discussion of errors can be found in Refs.~\cite{JE01b} and
\cite{CJ02} respectively. In Fig.~\ref{fig:S=0} the $\pi N$ phase
shift and inelastity in the $S_{11}$ channel, as well as transition
cross sections are depicted.  Finally, in Fig.~\ref{fig:S=-1} we show
our results for several amplitudes and cross sections in the $I=0$,
$S=-1$ channel, compared to the work of Ref.~\cite{ORB02} where
$\Delta_m = \Delta_M=0$. As we see, the description is quite
satisfactory.



Scattering lengths for the are given in Tab.~\ref{tab1}, compared
experiment and other determinations for the $I=1/2$, $S=0$ and $I=0$,
$S=-1$ channels. Errors are asigned by propagating best fit parameter
uncertainties including possible correlations~\cite{JE01b,CJ02}.

\begin{table}
\begin{tabular}{lrrrrrr}
\hline
  &  \tablehead{1}{r}{b}{$I$}
  & \tablehead{1}{r}{b}{$S$}
  & \tablehead{1}{r}{b}{Re a (fm) }
  & \tablehead{1}{r}{b}{Im a (fm) }
  & \tablehead{1}{r}{b}{Re a (fm) }
  & \tablehead{1}{r}{b}{Im a (fm) }   \\
\hline
$\pi N$ & 1/2 & 0 & 0.179(4) & 0 & 0.252(6)\tablenote{Experiment
Ref.\cite{Sch99}} & 0 \\
                       & & & & & 0.19(5)\tablenote{HBChPT
                       Ref.\cite{ej00c}} & 0 \\
$\eta N$ & 1/2 & 0  & 0.772(5) & 0.217(3) & 0.68\tablenote{Potential
model Ref.\cite{AS95}}   &
  0.24  \\
$ K \Lambda $ & 1/2 & 0  & 0.0547(5) & 0.032(4) &     &   \\
\hline
$ \pi \Sigma $ & 0   & -1 & 1.10(6) & 0 &     &  \\
$ \bar K N $  & 0   & -1 & -1.20(9) &  1.29(9) &
-1.71\tablenote{Experiment, Ref.~\cite{Ma81}}    & 0.68 \\
$\eta \Lambda$ & 0   & -1 & 0.50(5) &  0.27(1) &    &  \\
\hline
\end{tabular}
\caption{Real and imaginary parts of scattering lengths (in fm) for
best fit BSE results in several isospin and strangeness elastic
channels.}
\label{tab1}
\end{table}

\subsection{The static limit} 

In the static limit, baryons behave like fixed sources, and
consequently the two particle problem should reduce to a potential
scattering problem (in our case of meson-baryon scattering it would
correspond to a Klein-Gordon equation with a spin-dependent
potential). It is a known fact that the BSE has difficulties in
reproducing this heavy-light limit in certain situations (ladder
approximation to one boson exchange~\cite{gross82}). The remedy to
this situation is to make use of the Gross Equation, which essentially
means putting the would-be heavy particle on the mass shell from the
very beginning.  In Chiral Perturbation Theory this is also a tricky
point in the relativistic formulation, since dimensional
regularization does not lead to decoupling of heavy particles, which
has only been solved after a clever choice of renormalization scheme,
the so-called infrared regularization~\cite{bl01}.

The heavy baryon expansion may be taken by
making the baryon masses $\hat M\to \infty$ but keeping the meson
masses, $\hat m$, and  the meson momentum, $q$, finite and considering 
baryon mass splittings as higher order effects,
see e.g. \cite{Pich95}, 
\begin{eqnarray}
\hat M &=& M_B + \Delta \hat M 
\end{eqnarray} 
with $M_B \to \infty$ the common mass of the baryon octet which is
proportional to the identity matrix. Accordingly, in the $\pi N$
elastic channel we take
\begin{eqnarray}
\sqrt{s} = E+\omega = M_N + \omega + \frac{\omega^2-m_\pi^2}{2M_N} +
\cdots
\end{eqnarray}
where $M_N = M_B + \Delta M_N $.  Following Ref.~\cite{JE01b} in the
static limit one obtains from Eqs.~(\ref{eq:deff0})
and~(\ref{eq:t-1dirac}) ($f(\omega) \to -t(s)/(4\pi)$ )
\begin{eqnarray}
 f (\omega )^{-1} &=& 8 \pi \left[ \bar K_{\hat m} (\omega) +
\frac1{16 \pi^2} \ln \frac{\hat M^2}{\hat m^2} (\hat m-\omega)+ \hat M
J_{\hat m,\hat M}^0 +\frac{\Delta_{\hat m \hat M}^0}{4 \hat M} \right]
\nonumber \\ &-& \frac{4\pi}\omega\left\{ \Delta^0_{\hat m} - \left[
\frac2{f^2} D + \frac1{f^4}D \Delta^0_{\hat m} D \right]^{-1} \right\}
\label{eq:static}
\end{eqnarray} 
with the heavy baryon one-loop integral
\begin{eqnarray}
K_{\hat m} (\omega)= \frac1{ \rm i} \int { d^4 q \over (2\pi)^4 } {1\over q^2 - \hat
 m^2}{1\over \omega - v \cdot q } 
\end{eqnarray} 
and $\bar K_{\hat m} (\omega) = K_{\hat m} (\omega) -K_{\hat m} (m) -
(\omega-m) K'_{\hat m} (m)$ and the heavy baryon approximation of the
subtraction constants is denoted by the superscript
$0$. Eq.~(\ref{eq:static}) corresponds, as it should, to a one
particle scattering problem, fulfilling the coupled channel unitarity
condition
\begin{eqnarray} 
{\rm Im}  f (\omega)^{-1} = -\sqrt{\omega^2-\hat m^2} \theta (
\omega - \hat m) 
\end{eqnarray} 
The pole in Eq.~(\ref{eq:static}) for the inverse amplitude is a
static limit reminiscent from the baryonic Adler zero, $\sqrt{s}-\hat
M=0$, of the lowest order potential.  The constant combination
appearing in the inverse amplitude, Eq.~(\ref{eq:static}), $ \hat M
J_{\hat m\hat M}^0 + \Delta_{\hat m \hat M}^0 /4 \hat M$ should go to
some definite value in the static limit, $M \to \infty$. In case it
would diverge, the scattering amplitude would become
trivial. Numerical estimates in Ref.~\cite{JE01b} suggest that the
dangerous combination is not unnaturally large since
\begin{eqnarray} 
 \hat M J_{\hat m\hat M}^0 + \Delta_{\hat m \hat M}^0 /4 \hat M \sim
 \hat m 
\end{eqnarray}

\subsection{Chiral and heavy baryon expansions at threshold} 

If, in addition to a heavy baryon expansion, a chiral expansion in
powers of $1/f^2$ is carried out, one should recover in this double
expansion some form of the results found in
Ref.~\cite{Mo98,fms98,ej00c} within HBChPT for the elastic $\pi N$
scattering amplitude.  The results of Ref.~\cite{JE01b} show that the
matching to HBChPT is indeed possible, although there is an ambiguity
related to the absence of a left hand cut in our BSE amplitude,
Eq.~(\ref{eq:t-1dirac}), which is approximated with a polynomial in
the scattering region.  The net result is that there is a large degree
of redundancy in the BSE parameters, which hardly impose any practical
constraints on them\footnote{In this regard one should mention that
matching of unitarized amplitudes to HBChPT does not always work due
to a lack of convergence in HBChPT. See the discussion in
Refs.~\cite{JE01,pn00,ej00c}.}.

Moreover, a simple expansion in $1/f^2$ of the BSE amplitude keeping
the baryon masses does not converge well as one expects from the fact
that in the Born approximation the matrix elements $\pi N \to \eta N$,
$\eta N \to \eta N$ vanish, although in the full amplitude the
$N(1535)$ resonance is developed indicating strong rescattering
effects (see also below). For instance, taking $\pi N$ scattering
length for $I=1/2'$ obtained from a best fit to the data as an example one
gets~\cite{JE01b}
\begin{eqnarray}
a_{\pi N}^{I=1/2} = \underbrace{0.22}_{1/f^2} +
\underbrace{\overbrace{0.22}^{\pi N}- \overbrace{1.06}^{K
\Lambda}+\overbrace{0.18}^{K \Sigma}}_{1/f^4} + \dots = 0.179 \,~ {\rm
fm}
\end{eqnarray}
Note that the first two contributions leads to a value $-0.43 {\rm fm}
$.

\subsection{Resonances as poles in the second Riemann-sheet} 

The unitarity condition for the inverse amplitude, expressed as a
discontinuity equation reads~\cite{JE01b}
\begin{eqnarray}
{\rm Disc} \, [t^{-1} (s)] \equiv t_I^{-1} (s+{\rm i}\epsilon
)-t_I^{-1} (s-{\rm i}\epsilon ) = 2 {\rm i} \rho(s) \qquad s > (m+M)^2 
\end{eqnarray} 
with real $s$, where the phase space function
\begin{eqnarray}
\rho(s) = {\lambda^{1/2} (s,m_\pi,M_N) \over 16 \pi s } \times 
{ ( \sqrt{s}+M_N)^2 -m_\pi^2  \over 2 \sqrt{s} }
\end{eqnarray} 
has been introduced, understanding that $\rho(s)$ is a function of the
real variable $s$.  Then, analytically continuing the phase space
function to all complex plane, 
one finds the amplitude in the second Riemann sheet,
\begin{equation} 
t_{II}^{-1} (z) = t_I^{-1} (z)-2 {\rm i} \rho(z)  
\end{equation} 
Masses and widths computed as poles in the second Riemann sheet $ z_R
= M_R^2 - {\rm i} M_R \Gamma_R $, can be looked up at
Tab.~\ref{tab2}. The quoted theoretical errors are obtained
propagating the uncertainties in the best fit parametrs.

It is striking that the unobserved resonance has also been obtained by
the authors of Ref.~\cite{Jido:2002yz} although with $M=1390$ MeV and
$\Gamma=66$ MeV. Finally, one finds~\cite{CJ02} that the resonances
{\it are not} of the Breit-Wigner form, so that the simple relation
between residues and branching ratios does not hold\footnote{In the
one channel case the Breit-Wigner form implies a direct relation
between the residue at the pole and the imaginary part of the
pole. This relation only holds in the limit of sharp
resonances.}. Instead, an extrapolation to the real axis is required
to extract branching ratios as provided in the PDG~\cite{pdg02} for
which often Breit-Wigner or other particular forms are assumed. In
this regard, it would be desirable that future editions of the PDG
would also incorporate the residues.

\begin{table}
\begin{tabular}{lrrrrrrr}
\hline
    \tablehead{1}{r}{b}{$I$}
  & \tablehead{1}{r}{b}{$S$}
  & \tablehead{1}{r}{b}{$M_R$(MeV)}
  & \tablehead{1}{r}{b}{$\Gamma_R$(MeV)}
  & \tablehead{1}{r}{b}{PDG}
  & \tablehead{1}{r}{b}{$M_R$(MeV)}
  & \tablehead{1}{r}{b}{$\Gamma_R$(MeV)}   \\
\hline
1/2 & 0  & 1497(1) & 83(1) & N(1535) & 1505(10)    & 170(80)  \\
1/2 & 0  & 1684(1) & 194(1) & N(1660)  & 1660(20)    & 160(10)  \\
\hline
0   & -1 & 1368(12) & 250(23) & Unknown  &     &  \\
0   & -1 & 1443(3) &  50(7) & $\Lambda$ (1405) & 1407(4)    & 50(2) \\
0   & -1 & 1678(1) &  29(2) & $\Lambda$ (1670) & 1670(10)    & 35(12)
    \\
0   & -1 &   &   &     & 1673(2)\tablenote{Experiment. Ref.~\cite{Ma02}}& 23(6)  \\
\hline
\end{tabular}
\caption{Resonance Masses and Widths in MeV found for the best fit BSE 
calculation, compared with PDG data~\cite{pdg02}.}
\label{tab2}
\end{table}

\begin{theacknowledgments}
E.R.A. thanks the organizers of the workshop for the invitation and
the kind atmosphere provided in Coimbra. We warmly thank E. Oset and
A. Ramos for useful discussions. This research was supported by DGES
under contracts BFM2000-1326 and PB98-1367 and by the Junta de
Andalucia.
\end{theacknowledgments}


\bibliographystyle{aipproc}   



\end{document}